\documentclass[preprint,rmp,aps]{revtex4}

\usepackage{graphics}
\usepackage{epsfig}

\begin{document}

\title{ Emergence of Thermodynamics from Darwinian Dynamics }

\author{ P. Ao }

\address{ Department of Mechanical Engineering and Department of
Physics \\ University of Washington, Seattle, WA 98195, USA }





\begin{abstract}
Darwinian dynamics is manifestly stochastic and nonconservative,
but has a profound connection to conservative dynamics in physics.
In this short presentation the main ideas and logical steps
leading to thermodynamics from Darwinian dynamics are discussed in
a quantitative manner. It suggests that the truth of the second
law of thermodynamics lies in the fact that stochasticity or
probability is essential to describe Nature.  \\
{\ }  
{\ } \\
Presented at Keenan Symposium, {\it Meeting the Entropy Challenge},
October 4-5, 2007. MIT, Boston, USA.
To appears in Proceedings of Keenan Symposium, {\it Meeting the
Entropy Challenge}, October 4-5, 2007. MIT, Boston, USA. AIP Conference Proceedings
Series.

\end{abstract}

\maketitle

It's kind of awkward to do the talking seated with face away
from the big screen. May I use the podium? (pause)
This morning we heard lots of very nice talks.
They have, I think, laid out a good foundation for my few minutes.
I am grateful for the organizers
for giving me the opportunity and honor to speak to the
distinguished audience. What I'm going to tell you are a few insights
that I got from biology. I will take Darwin and Wallace seriously.

In order to understand thermodynamics, a kind of a static property
with no time dependence, the proper way should be that started from a
real dynamical structure. What I have learned in biology is that there is
already one, which is more than 100 years old, and was
from Darwin and Wallace \cite{darwin1858}. Why should anybody do this
demonstration? It is the focus of the present conference,
the foundation of thermodynamics and statistical mechanics, which have
been under vigorous investigation for more than 100 years
\cite{nicolis,mit,xing,emch,gemmer,qian,ao2005a,attard}--J.H. Keenan
had thought deeply
into this problem \cite{keenan}. It is fair to state that no
consensus has been reached.

There are two types of important questions in this perspective.
One is of course on the foundation. For example,
three major questions have been formulated recently \cite{uffink}:  \\
 1) In what sense can thermodynamics be reduced to statistical mechanics? \\
 2) How can one derive equations that are not time-reversal invariant
    from a time-reversal invariant dynamics? \\
 3) How to provide a theoretical basis for the "approach to equilibrium"
     or irreversible processes? \\
With the aid of the evolutionary dynamics of Darwin and Wallace,
Darwinian dynamics in short,
my answer to 1) is that thermodynamics and statistical mechanics are
equivalent. They are the two sides of same coin. The answer to 2)
is that additional primitive concepts, such as probability and
stochasticity, have to be introduced. Not time-reversal
invariant dynamics,  while consistent with, is logically independent from
a time-reversal invariant one. Thus, the answer to 3)
is that Darwinian dynamics provides a natural theoretical basis.

Another type of questions is from a dazzling on the enormous success of
thermodynamics and statistical mechanics. It has been wondering that \cite{fine}, \\
 4) Such success would be irrelevant to inference and decision-making; and \\
 5) were assured by unstated methodological practices of censoring data
      and selective applying arguments; or  \\
 6) is a result of extraordinary good fortune. \\
It is evident that 4) is not true.
The connection between statistical mechanics and inference
has been well established \cite{jaynes}, two sides of our effort to
understand Nature. 5) is not true either.
Physicists have not used any arbitrary and selective methodology.
Nevertheless, 6) contains certain truth: It has been shown \cite{ao2005b} that
Darwinian dynamics which is adaptive has a remarkable connection to
conservative dynamics. Such a relation between two extreme
theoretical frameworks demonstrates the unity of sciences.

You may ask then, what is Darwinian dynamics? Word equation of the
evolutionary dynamics discovered by Darwin and Wallace
\cite{darwin1858} is:

\begin{center}
  {\it Evolution by Variation and Selection. } {\ } {\ }  {\ } {\ }  {\ } {\ }
 {\ } {\ } (0)
\end{center}

For nearly 150 years since its conception, this word equation has
been applied to all levels of biology \cite{wilson}.
There is no exception yet for its validity.
While universal, it is not in the mathematical form as
expressions in thermodynamics and statistical mechanics. This word
equation needs to be expressed in terms of proper mathematical
equations. Two further fundamental concepts in biology are needed for this purpose.
The first is the so-called fundamental theorem of natural selection
(FTNS) \cite{fisher} which links the variation to the adaptation and optimization.
In the hindsight, it appears to be equivalent to
fluctuation-dissipation theorem in nonequilibrium physics \cite{ao2005b,kaneko}.
The second is the existence of adaptive landscape \cite{wright}.
It is equivalent to the assertion of the existence of potential
function or Hamiltonian in an evolutionary process \cite{ao2005b}.

Intuitively it is evident that evolution is about successive
processes: Quantities at a later stage are related to their values
at its earlier stage under both predictable (deterministic) and
unpredictable (stochastic) constraints. For example, the world
population of humans in next 20 years will be surely related to
its current one. Hence, the genetic frequency, the probability in
the population, of a given form of gene (allele) in the next
generation is related to its present value. Here sexual conducts
and other reproduction behaviors are treated as means to realize
the variation and selection for evolution. We may denote those
genetic frequencies as ${\bf q}$ with $n$ components denoting all
possible alleles. Thus ${\bf q}^{\tau} = (q_1, q_2, ... , q_n)$ is
a vector (Here $\tau$ denotes the transpose). There are huge
amount of human traits related to genetics (or genes): height,
skin color, size of eye ball, faster runner, gene for liver
cancer, gene for smartness, ... . The number $n$ is then large: it
could be as large, and likely larger, as the number of genes in
human genome, which is about 20000, if one simply assigns one
allele or a trait to one gene without any combinatory
consideration. This number is far larger than the number of
chemical elements, which is about 100, and than the number of
elementary particles, about 30. With a suitable choice of time
scale equivalent to an averaging over many generations, the
incremental rate in such an evolutionary process may be
represented by a time derivative, ${\dot {\bf q}} = d {\bf q}/dt$.
The deterministic constraint at a given time may be represented by
a deterministic force ${\bf f}({\bf q})$. For example, there is
a high confidence to predict the eye color of a child based on the
information from his/her parents, but the smartness of an
offspring is not so strongly correlated to that of the parents.
The random constraint, the unknown and/or irrelevant force, is
approximated by a Gaussian-white noise term ${\bf \zeta}({\bf q},
t)$, with zero mean, $\langle {\bf \zeta} \rangle = 0$ and the $n
\times n$ matrix $D$: $ \langle {\bf \zeta}({\bf q}, t')
{{\bf \zeta}}^{\tau}({\bf q}, t) \rangle = 2 D({\bf q}) \theta
\delta (t-t') $. Here the factor 2 is a convention and $\theta$ is
a positive numerical constant reserved for the role of temperature
in physical sciences. $\delta(t)$ is the Dirac delta function.
With these notations we are ready to transform the word equation
into a precise mathematical equation, which reads
\begin{eqnarray}
 {\dot {\bf q}} & = & {\bf f}( {\bf q} ) + {\bf \zeta}({\bf q}, t) \; .
\end{eqnarray}
However, an immediate question arises: while we may represent the
variation in evolution by the matrix $D$, where is
Wright's adaptive landscape and the corresponding potential
function?

The deterministic force ${\bf f}({\bf q})$ in general cannot be related to
a potential function in a straightforward way,
because ${\bf f}({\bf q}) \neq  D({\bf q}) \nabla \phi({\bf q})$.
Here $\nabla = (\partial/\partial
q_1, \partial/\partial q_2, ... , \partial/\partial q_n)^{\tau} $
is the gradient operation in the phase space formed by ${\bf q}$,
and $ \phi({\bf q})$ is a scalar function.
A nonequilibrium process typically has further
five qualitative characteristics or difficulties:  \\
 i) dissipative,  $\nabla \cdot {\bf f}({\bf q}) \neq 0 $;  \\
 ii) asymmetric, $\partial f_j({\bf q})/ \partial q_i \neq
  \partial f_i({\bf q})/ \partial q_j$ for at least one pair of indices of $i,j$; \\
 iii) nonlinear, $ {\bf f}(\theta {\bf q}) \neq \theta \, {\bf f}({\bf q})$; and  \\
 iv) stochastic with multiplicative noise, $D({\bf q})$ depending on the state variable
  ${\bf q}$.  \\
 v) Possibly singular, that is, $\det( D({\bf q}) ) = 0$. \\

During the study of the robustness of the genetic switch in a
living organism \cite{zhu2004a,zhu2004b} a constructive method was
discovered to overcome this difficulty \cite{ao2004,kat,yin}:
Eq.(1) can be transformed into the following stochastic differential equation,
\begin{equation} \label{normal}
 [ A({\bf q}) + T({\bf q})] \dot{{\bf q}} = \nabla \phi({\bf q}) + \xi({\bf q},t) \; ,
\end{equation}
with $ \langle {\bf \xi}({\bf q}, t') {{\bf \xi}}^{\tau}({\bf q},
t) \rangle = 2 A({\bf q}) \theta \delta (t-t') $ and $T = - T^{\tau}$. Here the
matrices $A,T$ are determined from $D$ and ${\bf f}$ by two matrix equations:
$
 [ A({\bf q}) + T({\bf q}) ] D({\bf q}) [ A({\bf q}) - T({\bf q}) ]
   = A({\bf q})
$
and
$
 \nabla \times [ [A({\bf q}) + T({\bf q}) ] {\bf f}({\bf q}) ] = 0
$.
With this construction done, Wright's adaptive landscape can be easily identified with the
scalor function $\phi$.
Darwinian dynamics may then be summarized into three general
laws in the Table I \cite{ao2005b} .

{\ }

\noindent
\begin{tabular}{|c|c|c|c|}
  \hline
    & mathematical expressions & alternative names &  comments   \\
  \hline
 First Law                     &
  $  \{ {\bf q}  \}
       \rightarrow  \{ {\bf q}_{attractor} \} $  &
    law of Aristotle           &
   determinism               \\
 Second Law                    &
  $ [ A({\bf q}) + T ({\bf q}) ] \dot{\bf q}
     = \nabla \phi ({\bf q}) + {\bf \xi}({\bf q}, t) $ &
    law of Darwin {\&} Wallace  &
       stochasticity      \\
 F-Theorem                     &
  $ \langle {\bf \xi}({\bf q}, t)
          {\bf \xi}^{\tau} ({\bf q}, t') \rangle
    = 2 A({\bf q})  \theta \; \delta(t-t') $   &
    FTNS              &
    optimization    \\
 Third Law                     &
     $m$ $\rightarrow 0 $     &
    law of hierarchy           &
    multiple scales   \\
  \hline
\end{tabular}

{\ }

\noindent { Table I: Laws of Darwinian dynamics and the F-Theorem}

{\ }

With above general quantitative formulation we are ready to
move from biology to physics. We first point out two immediate
consequences of Darwinian dynamics.

Allowing the stochastic drive be negligible, "temperature" $\theta = 0$,
Eq.(2) becomes
\begin{equation}
  [ A({\bf q}) + T ({\bf q}) ] \dot{\bf q}
     = \nabla \phi ({\bf q})  \; .
\end{equation}
Because the ascendant matrix $A$ is non-negative and $T$ is anti-symmetric,
the system will approach the nearest attractor determined by its initial
condition, and stay there if already there. Specifically, Eq.(3)
leads to
\begin{equation}
  \dot{ \bf q} \cdot \nabla \phi ({\bf q}) \geq 0 \; .
\end{equation}
This equation implies that the deterministic dynamics cannot
decrease the evolutionary potential $\phi$.
If the ascendant matrix $A$ is positive definite,
the potential of the system always increases. Hence, the first law
clearly states that the system has the ability to find the local
adaptive landscape peak or an attractor represented by the Wright evolutionary
potential $\phi$, determined by the initial condition. However, the
shifting between different evolutionary peaks would become
impossible in this limit, because the transition probability
vanishes exponentially. We note that Eq.(11) implies that the
Wright evolutionary potential $\phi$ is a Lyapunov function.

Conservative Newtonian dynamics may be regarded as a further limit
of the above formulation: zero friction and zero noise limit, $ A =0 $.
Hence, from Eq.(3), Newtonian dynamics may be expressed as,
\begin{equation} \label{newton}
  T({\bf q}) \; \dot{{\bf q}} = \nabla  \phi({\bf q} )  \; ,
\end{equation}
which is the form similar to Hamilton equation. This suggests a profound
connection between biology and physics.
Here the value of potential function is evidently conserved during
the dynamics since  $\dot{{\bf q}} \cdot  \nabla \phi({\bf q} ) = 0 $.
The system moves along equal potential contours in
the adaptive landscape. This conservative behavior suggests that
the rate of approaching to equilibrium is associated with the
ascendant matrix $A$, not with the diffusion matrix $D$. There are
situations where the diffusion matrix is finite but the ascendant
matrix is zero, and thus the dynamics is conservative \cite{zhu2006}.

Darwinian dynamics can be described by a probabilistic
equation, a special form of Fokker-Planck equation. It has been
derived from Eq.(2) \cite{yin} that the equation for probability
distribution function $ \rho $ is
\begin{equation}
  \partial_t \rho({\bf q},t)
   = \nabla^{\tau} [ [ D({\bf q}) + Q({\bf q}) ] \theta \nabla
       - {\bf f}({\bf q}) ] \rho({\bf q}, t) \; ,
\end{equation}
with $ [ D({\bf q}) + Q({\bf q}) ] = [ A({\bf q}) + T({\bf q}) ]^{-1} $ and
$ {\bf f}({\bf q}) = [ D({\bf q}) + Q({\bf q}) ] \nabla \phi({\bf q}) $.
Its steady state distribution is of Boltzmann-Gibbs type, if it exists:
\begin{equation}
\rho({\bf q}, t=\infty) = \frac{1}{Z} \exp \left\{
        \frac{ \phi({\bf q}) }{ \theta } \right\} \; .
\end{equation}
Here ${Z} = \int \prod_{i=1}^{n} d{q_i} \exp \left\{ {\psi({\bf
q}) } /{ \theta } \right\} $, the partition function.

We note that three independent parts in a general dynamics are suggested by
Eq.(6) and (7): The non-conservative dynamics associated with diffusion matrix $D$,
the conservative dynamics with $Q$, and the potential function with $\phi$.
Thus, this is a remarkable synthesis between near and far from equilibrium processes,
and conservative and non-conservative dynamics.
This feature is also true for discrete master equations, though usually a different
interpretation is provided \cite{zia}.

I hope I have made it logically clear that the Boltzmann-Gibbs distribution
follows in a natural and consistent manner from Darwinian dynamics.
Starting from this distribution it is a relative straightforward
procedure to deduce rigorously the second law of thermodynamics \cite{ao2005a,ma}.
I will not waste your time here by going through such process.
Instead, I should point out that
further new results on stochastic equalities can be obtained from Darwinian dynamics,
such as a generalized Einstein relation,
a free energy dynamical equality in the absence of
detailed balance \cite{ao2005a}.
I should also point out that among the three universal
dynamics known to us, Darwinian dynamics, relativity, and
quantum mechanics, only Darwinian dynamics is non-conservative.
This may be the reason for the natural emergence of thermodynamics.
Thank you very much for your attention.

{\ }

This work was supported in part by USA NIH grant under HG002894.



%
%

\end{document}